\documentclass[aps,prd,preprint,a4paper,showpacs,nofootinbib,superscriptaddress]{revtex4-2}
\usepackage{bm}
\usepackage{indentfirst}
\usepackage{amsmath}
\usepackage{graphicx}
\usepackage{float}
\usepackage{amssymb}
\usepackage{subfigure}
\usepackage{amssymb}
\usepackage{hyperref}
\usepackage{epstopdf}
\usepackage[section]{placeins}
\usepackage{soul}

\usepackage[utf8]{inputenc}
\hypersetup{
    colorlinks=true,
    linkcolor=red,
    citecolor=blue,
}
\usepackage{color}
\usepackage[T1]{fontenc}
\usepackage{txfonts}
\usepackage{orcidlink}

\usepackage{adjustbox,lipsum}

\newcommand{\Rmnum}[1]{\expandafter\@slowromancap\romannumeral #1@} 
\newcommand{\bq}{\begin{equation}}
\newcommand{\eq}{\end{equation}}
\newcommand{\bqn}{\begin{eqnarray}}
\newcommand{\eqn}{\end{eqnarray}}

\newcommand{\lb}{\label}

\makeatother

\begin{document}

\title{Spectral instability in modified P\"oschl-Teller effective potential triggered by deterministic and random perturbations}

\author{Shui-Fa Shen}
\affiliation{School of Intelligent Manufacturing, Zhejiang Guangsha Vocational and Technical University of Construction, 322100, Jinhua, Zhejiang, China}
\affiliation{Department of Scientific Research, Innovation and Training of Scientific and Pedagogical Staff, University of Economics and Pedagogy, Karshi 180100, Uzbekistan}

\author{Guan-Ru Li}
\affiliation{Faculdade de Engenharia de Guaratinguet\'a, Universidade Estadual Paulista, 12516-410, Guaratinguet\'a, SP, Brazil}

\author{Ramin G. Daghigh}
\affiliation{Natural Sciences Department, Metropolitan State University, Saint Paul, Minnesota, 55106, USA}

\author{Jodin C. Morey}
\affiliation{Le Moyne College, Syracuse, New York, 13214, USA}

\author{Michael D. Green}
\affiliation{Mathematics and Statistics Department, Metropolitan State University, Saint Paul, Minnesota, 55106, USA}

\author{Wei-Liang Qian}\email[E-mail: ]{wlqian@usp.br}
\affiliation{Escola de Engenharia de Lorena, Universidade de S\~ao Paulo, 12602-810, Lorena, SP, Brazil}
\affiliation{Faculdade de Engenharia de Guaratinguet\'a, Universidade Estadual Paulista, 12516-410, Guaratinguet\'a, SP, Brazil}
\affiliation{Center for Gravitation and Cosmology, College of Physical Science and Technology, Yangzhou University, Yangzhou 225009, China}

\author{Rui-Hong Yue}
\affiliation{Center for Gravitation and Cosmology, College of Physical Science and Technology, Yangzhou University, Yangzhou 225009, China}

\begin{abstract}
Owing to its substantial implications for black hole spectroscopy, spectral instability has attracted considerable attention in the literature.
While the emergence of such instability is attributed to the non-Hermitian nature of the gravitational system, it remains sensitive to various factors.
About the spatial scale of the metric deformation, spectral instability is particularly susceptible to ``ultraviolet'' metric perturbations.
In this work, we conduct a focused analysis of black hole spectral instability using the P\"oschl-Teller potential as a toy model.  
We investigate the dependence of the resulting spectral instability on the magnitude, spatial scale, and localization of deterministic and random perturbations in the effective potential of the wave equation, and discuss the underlying physical interpretations.
It is observed that small perturbations in the potential initially have a limited impact on the less damped black hole quasinormal modes with deviations typically around their unperturbed values, a phenomenon first derived by Skakala and Visser in a more restrictive context.
In the higher overtone region, the deviation propagates, amplifies, and eventually gives rise to spectral instability and, inclusively, bifurcation in the quasinormal mode spectrum. 
While deterministic perturbations give rise to a deformed but well-defined quasinormal spectrum, random perturbations lead to uncertainties in the resulting spectrum.
Nonetheless, the primary trend of the spectral instability remains consistent, being sensitive to both the strength and location of the perturbation.
However, we demonstrate that the observed spectral instability might be suppressed for perturbations that are physically appropriate.

\end{abstract}

\date{July 27th, 2025}

\maketitle

\newpage
\section{Introduction}\label{sec1}

Black holes are among the most intriguing concepts in theoretical physics. 
They exemplify gravity's properties at their extremity through an elegant and concise mathematical framework. 
The success of the ground-based LIGO and Virgo collaborations in capturing gravitational waves emanating from binary mergers~\cite{agr-LIGO-01, agr-LIGO-02, agr-LIGO-03, agr-LIGO-04}, as well as ongoing spaceborne projects, such as LISA~\cite{agr-LISA-01}, TianQin~\cite{agr-TianQin-01}, and Taiji~\cite{agr-Taiji-01}, marks the inauguration of a new epoch for observational astrophysics. 
In particular, the feasibility of direct observation of ringdown waveforms has also been explored~\cite{agr-TianQin-05}.

The black hole ringdown waveform is primarily determined by a linear combination of quasinormal modes (QNMs)~\cite{agr-qnm-review-02, agr-qnm-review-03}, whose study has sparked considerable interest. 
It is well-known~\cite{agr-qnm-21, agr-qnm-29} that, in the frequency space, QNMs can be attributed to the poles of the Green's function related to the master wave equation, whereas the late-time behavior is associated with the branch cut, which typically corresponds to an exponential decay followed by an inverse power tail in the time domain~\cite{agr-qnm-tail-01}. 
Along this line of thought, it has been understood in the literature that the unique characteristic of QNMs, determined solely by the spacetime properties surrounding the black hole, provides unambiguous information about the spacetime geometry near the event horizon.

However, to some extent, this concept has been challenged recently by the renewed discussion of spectral instability~\cite{agr-qnm-instability-02, agr-qnm-instability-03, agr-qnm-instability-11, agr-qnm-lq-03, agr-qnm-instability-07, spectral-instability-review-20}. 
Pioneered by Nollert and Price~\cite{agr-qnm-instability-02, agr-qnm-instability-03}, it was shown that even minor perturbations in the Regge-Wheeler potential, such as step functions, can significantly influence the higher overtone modes in the QNM spectrum, demonstrating an unexpected instability of the QNM spectrum against small-scale perturbations. 
This finding challenges the conventional assumption that a reasonable approximation of the effective potential is expected to result in only minimal deviation in the resulting QNMs. 
Further analytic and numerical studies~\cite{agr-qnm-instability-11, agr-qnm-lq-03} have indicated that even in the presence of discontinuities of a more moderate nature, the asymptotic behavior of the QNM spectrum could be non-perturbatively modified. 
Specifically, high-overtone modes might shift along the real frequency axis rather than ascending the imaginary frequency axis observed for most black hole metrics~\cite{agr-qnm-continued-fraction-12, agr-qnm-continued-fraction-23}. 
It was argued that this feature persists regardless of the discontinuity's distance from the horizon or its magnitude.
Generalizing these results, Jaramillo {\it et al.}~\cite{agr-qnm-instability-07} explored the topic in terms of the notion of spectral instability. 
Their analyses revealed that the boundary of the pseudospectrum moves closer to the real frequency axis, thus reinforcing the notion of a universal instability of high-overtone modes triggered by ``ultraviolet,'' meaning small-scale, perturbations.

Because in a real-world astrophysical context gravitational radiation sources such as black holes are hardly isolated objects, the above results have an immediate and substantial implication in black hole spectroscopy, which involves modeling of black hole waveforms and extracting the underlying metric parameters, inclusively involving the temporal profiles dominated by QNMs~\cite{agr-bh-spectroscopy-05, agr-bh-spectroscopy-06, agr-bh-spectroscopy-15, agr-bh-spectroscopy-18, agr-bh-spectroscopy-20, agr-bh-spectroscopy-36, agr-qnm-review-05, agr-BH-spectroscopy-review-04}. 
The fact that black holes are typically submerged and interact with the surrounding matter leads to deviations from the ideal symmetric metric.
As a result, the emitted gravitational waves of the underlying QNMs might differ substantially from those predicted for a pristine, isolated, compact object. 
In the literature, such phenomena have motivated the investigation into ``dirty'' black holes, as explored by several authors~\cite{agr-bh-thermodynamics-12, agr-qnm-33, agr-qnm-34, agr-qnm-54}, and opened new avenues in the study of black hole perturbation theory.
Moreover, the asymptotic modes that align almost parallel to the real axis are closely related to the intriguing concept of echoes, a late-stage ringing waveform first proposed by Cardoso {\it et al.}~\cite{agr-qnm-echoes-01, agr-qnm-echoes-review-01}. 
As potential observables, echoes might help distinguish different but otherwise similar gravitational systems via their distinct properties near the horizon. 
This idea has spurred many studies into echoes across various systems, encompassing exotic compact objects such as gravastars and wormholes. 
Like the late-time tail, echoes are also attributed to the analytic properties of Green's function, as analyzed by Mark {\it et al.}~\cite{agr-qnm-echoes-15}. 
In studies of Damour-Solodukhin type wormholes\cite{agr-wormhole-12}, Bueno {\it et al.}~\cite{agr-qnm-echoes-16} have investigated echoes by explicitly solving for specific frequencies at which the transition matrix becomes singular, providing further insights into the complex interplay of spacetime geometry and QNMs.
For the above scenarios, the perturbations to the metric are not necessarily located close to the event horizon.
In other words, the apparent deformation in the frequency-space QNM spectrum, and potentially the consequential time-domain waveforms, might not be entirely attributed to the distortion of the spacetime curvature near the compact object in question.

The related topic of spectral instability, echoes, greybody factor, and causality has been extensively explored in recent years by many authors~\cite{agr-qnm-instability-08, agr-qnm-instability-13, agr-qnm-instability-14, agr-qnm-instability-15, agr-qnm-instability-16, agr-qnm-instability-18, agr-qnm-instability-19, agr-qnm-instability-22, agr-qnm-instability-29, agr-qnm-echoes-20, agr-qnm-echoes-22, agr-qnm-instability-26, agr-qnm-instability-32, agr-qnm-instability-33, agr-qnm-instability-47, agr-qnm-instability-55, agr-qnm-instability-56, agr-qnm-echoes-45, agr-qnm-instability-60, agr-qnm-instability-61, agr-qnm-instability-63, agr-qnm-instability-65, agr-qnm-Regge-14, agr-qnm-instability-72, agr-qnm-instability-83} (see~\cite{spectral-instability-review-20} for a more comprehensive list of references).
Notably, Cheung {\it et al.}~\cite{agr-qnm-instability-15} pointed out that even the fundamental mode can be destabilized under generic perturbations.
By introducing a small perturbation to the Regge-Wheeler effective potential, it was shown~\cite{agr-qnm-instability-15} that the fundamental mode undergoes an outward spiral while the deviation's magnitude increases.
Subsequently, its position is overtaken by a new mode distinct from the first few low-lying modes. 
The results have been ascertained using different shapes for a small bump in the potential and observed in a toy model constituted by two disjointed rectangular potential barriers~\cite{agr-qnm-instability-15, agr-qnm-instability-32} and derived from an analytic account~\cite{agr-qnm-instability-56, agr-qnm-instability-58, agr-qnm-instability-55, agr-qnm-instability-50}.
Further analysis revealed that the fundamental mode can remain stable, depending sensitively on the specific background metric and perturbations~\cite{agr-qnm-instability-55}.
The observed interplay between the fundamental mode and echo modes can also be properly assessed~\cite{agr-qnm-instability-65}.
Although it has been argued that spectral instability might not significantly impact the time-domain waveform~\cite{agr-qnm-instability-16}, these observations potentially undermine the feasibility of black hole spectroscopy as the fundamental mode is shown to be subjected to spectral instability.
Besides the scale of the perturbation, the above findings indicate that its location is also relevant.
In this regard, a systematic analysis of the change in the spectrum in relation to different characteristics of perturbations in the potential is still lacking in the literature.

The present study is motivated by the above considerations.
Our analysis focuses on the sensitivity of the QNM spectrum against the magnitude, spatial scale, and localization of the perturbations in the effective potentials.
To this end, we consider the theoretical setup that the study of black hole perturbations can be simplified by exploring the spatial part of the master equation~\cite{agr-qnm-review-03},
\begin{eqnarray}
\frac{\partial^2}{\partial t^2}\Psi(t, r_*)+\left(-\frac{\partial^2}{\partial r_*^2}+V_\mathrm{eff}\right)\Psi(t, r_*)=0 ,
\label{master_eq_ns}
\end{eqnarray}
where  $r_*$ is the tortoise coordinate, and $V_\mathrm{eff}$ is the effective potential determined by the spacetime metric, spin ${\bar{s}}$, and angular momentum $\ell$ of the perturbation.
For instance, the P\"ochl-Teller potential is given by
\begin{equation}
V_\mathrm{PT}\equiv {V_0}~{\text{sech}}^2\left(\frac{r_{*}}{b}\right),
\label{potential_PT}
\end{equation}
where $V_0$ and $b$ are two parameters governing the shape of the potential.

The quasinormal frequencies can be obtained by evaluating the zeros of the Wronskian,
\begin{eqnarray}
W(\omega)\equiv W(g,f)=g(\omega,r_*)f'(\omega, r_*)-f(\omega,r_*)g'(\omega,r_*) ,
\label{pt_Wronskian}
\end{eqnarray}
where $'\equiv d/dr_*$, and $f$ and $g$ are the solutions of the corresponding homogeneous equation of Eq.~\eqref{master_eq_ns} in the frequency domain~\cite{agr-qnm-review-02},
\begin{eqnarray}
\left[-\omega^2-\frac{d^2}{dr_*^2}+V_\mathrm{eff}\right]\widetilde{\Psi}(\omega, r_*)=0 ,
\label{pt_homo_eq}
\end{eqnarray}
with appropriate boundary conditions, namely,
\begin{eqnarray}
\begin{array}{cc}
f(\omega, r_*)\sim e^{-i\omega r_*}   ~~~~ &  r_*\to -\infty  ,\cr\\
g(\omega, r_*)\sim e^{i\omega r_*}    ~~~~ &  r_*\to +\infty  .
\end{array} 
\label{pt_boundary}
\end{eqnarray}
As discussed above, the present paper aims to analyze the impact of different forms of metric perturbations on the resulting deformation of the QNM spectrum.  
In this study, for mathematical simplicity, instead of deriving the perturbed effective potential from physically relevant metric perturbations, we introduce the perturbation directly into the effective potential $V_\mathrm{eff}$.  
Following a practice commonly adopted by other authors~\cite{agr-qnm-instability-02, agr-qnm-instability-07, agr-qnm-instability-15}, the main motivation for this approach is mathematical simplicity, which enables us to focus on the essential features of the spectral instability.  
We note that the physical relevance of such metric perturbations is an important issue that has also been highlighted by other authors~\cite{agr-qnm-instability-47}.

The remainder of the paper is organized as follows.
In Sec.~\ref{sec2}, we start our analysis with deterministic perturbations in the P\"oschl-Teller potential, from which we discuss the resulting impact on the QNM spectrum and spectral instability.
The dependence of the deformation in the spectrum on the magnitude and frequency of the perturbations is explored.
In Sec.~\ref{sec3}, we further analyze the case where the perturbation in the effective potential possesses an arbitrary shape.
Subsequently, we elaborate on the scenario in which a small-scale perturbation is introduced to the background black hole metric and investigate the effect of its location in Sec.~\ref{sec4}.
The last section includes further discussions and concluding remarks.
In this study, numerical calculations are carried out using the matrix method, which has been verified by repeating the calculations with different grid sizes and precisions.

\section{Spectral instability triggered by deterministic perturbations}\label{sec2}

We consider the perturbed P\"oschl-Teller potential of the following form
\bqn
V_\mathrm{eff} = V_\mathrm{PT} + V_\mathrm{pert},
\lb{Veff_MPT}
\eqn
where we assume the parameters $V_0=b=5$ for the P\"oschl-Teller potential Eq.~\eqref{potential_PT}.
The quasinormal frequencies of the original P\"ochl-Teller potential are given by~\cite{agr-qnm-Poschl-Teller-02}
\bqn
\omega_n^\mathrm{PT}=\sqrt{V_{0}-\frac{1}{4b^2}}-i\frac{1}{b}\left(n+\frac{1}{2}\right),
\lb{qnm_PT}
\eqn
which implies that $\mathrm{Re}\omega_n \sim 2.23383$ and a constant distance of $0.2$ along the imaginary axis between successive modes, for the given parameters.
For the deterministic case, the perturbation takes the form
\bqn
V_\mathrm{pert} = V_\mathrm{pert}^\mathrm{det} = \epsilon \sin(k \pi r_*) ,\label{V_per_der}
\eqn
where the magnitude $\epsilon$ and wave number $k$ will be tuned in our calculations, the compatified coordinante
\begin{eqnarray}
x = \tanh(r_*) .\label{tcPT}   
\end{eqnarray}
Regarding the specific form of Eq.~\eqref{V_per_der}, some comments are in order.
As pointed out in~\cite{agr-qnm-instability-07}, the QNM spectrum is sensitive to small-scale perturbations.
The parameterization of Eq.~\eqref{V_per_der} features an explicit dependence on the perturbation's spatial scale via the parameter $k$.
Secondly, when transformed from the compactified coordinate $x$ to the tortoise coordinate $r_*$, this perturbation is exponentially suppressed at the bound $x=\pm 1$.
Specifically, as $r_*\to \pm\infty$ we have $x\sim \pm \left[1-2\exp(- 2|r_*|) \right]$, and subsequently, 
\begin{equation}
|V_\mathrm{pert}^\mathrm{det}|\sim 2\epsilon k\pi \exp(- 2|r_*|) \ll V_0\exp(-2|r_*|/b) \sim V_\mathrm{PT} ,\nonumber
\end{equation}
given $b \ge 1$ and $\epsilon k \ll V_0$.
In other words, this perturbation is minor in size compared to the original black hole potential.

\begin{figure}[ht]
\centerline{
\hspace{0.3cm}
\includegraphics[height=0.35\textwidth]{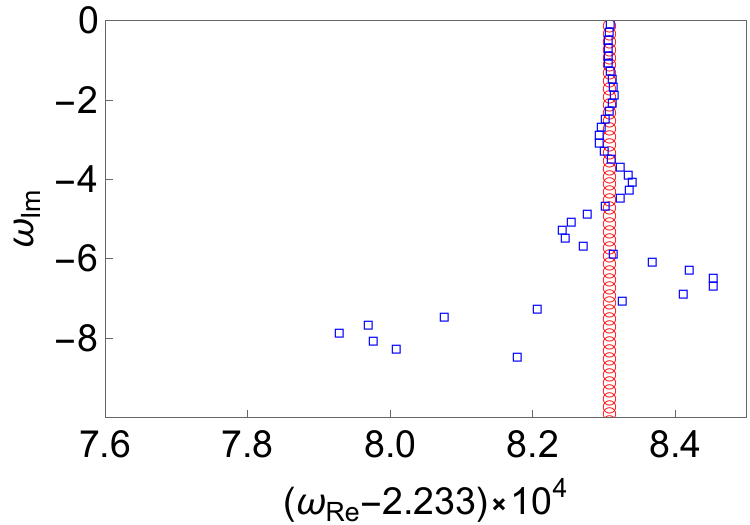}
\hspace{0.2cm}
\includegraphics[height=0.35\textwidth]{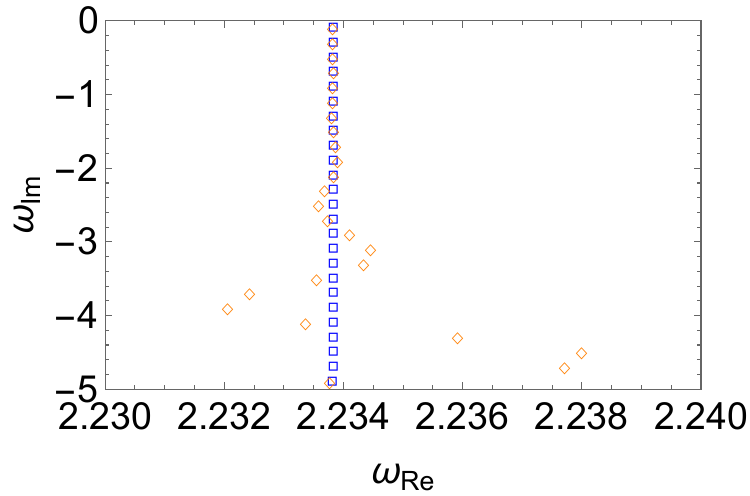}
}
\centerline{
\includegraphics[height=0.35\textwidth]{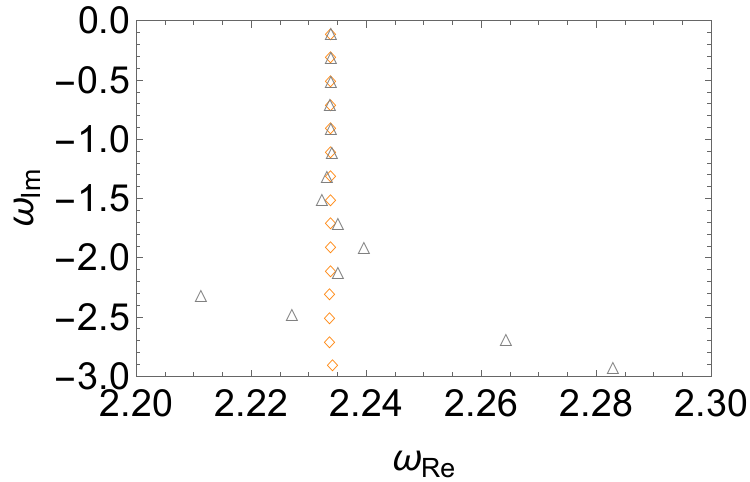}
\includegraphics[height=0.35\textwidth]{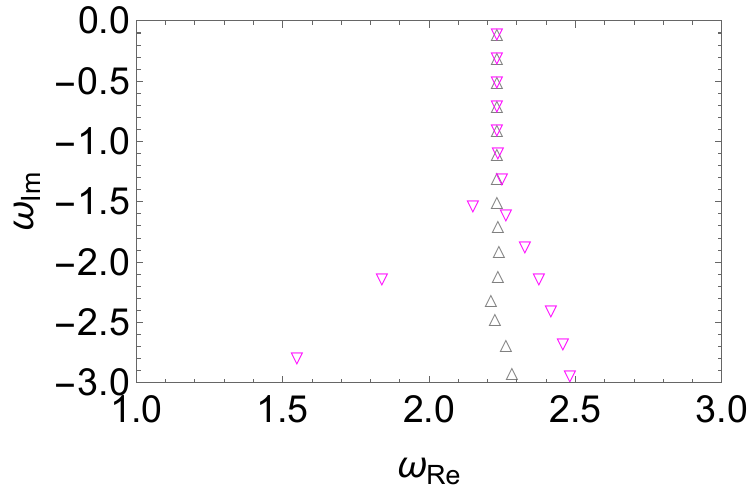}
}
\centerline{
\includegraphics[height=0.35\textwidth]{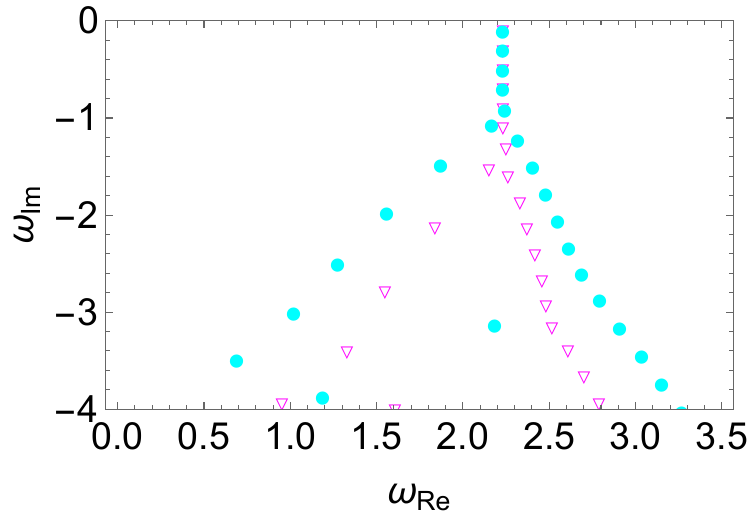}
\includegraphics[height=0.35\textwidth]{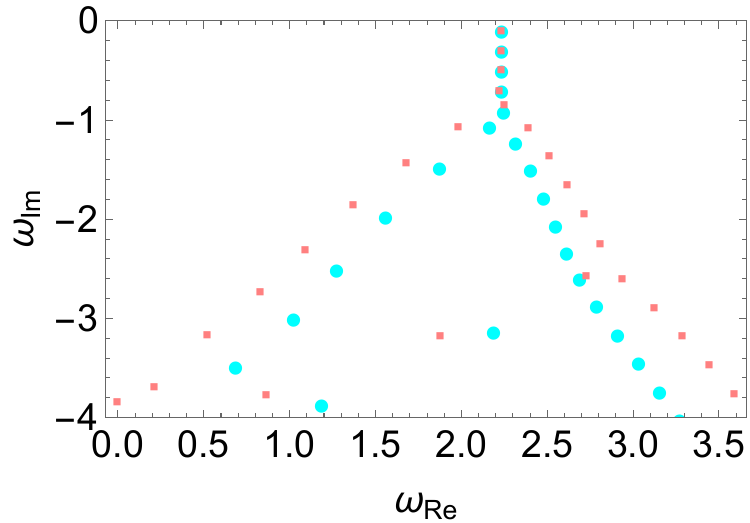}
}
\renewcommand{\figurename}{Fig.}
\vspace{-0.5cm}
\caption{The spectral instability in the perturbed P\"oschl-Teller effective potential Eq.~\eqref{Veff_MPT} triggered by deterministic metric perturbation Eq.~\eqref{V_per_der}, where one assumes $V_0=b=5$ and $\epsilon=0.001$.
From left to right and top to bottom, we compare in pairs the resulting QNM spectra for different metric perturbations using the wave numbers $k=0$ (unperturbed metric, empty red circles), $1.0$ (empty blue squares), $2.0$ (empty orange diamonds), $3.0$ (empty gray triangles), $4.0$ (empty magenta flipped triangles), $5.0$ (filled cyan circles), and $6.0$ (filled pink squares).
The numerical calculations are carried out using the matrix method, and the results have been verified using different grid sizes.}
\label{fig_PT_det_frequency}
\end{figure}

The calculations are performed using the matrix method~\cite{agr-qnm-lq-matrix-02}.  
This approach reformulates the master equation for QNMs into an algebraic one for complex eigenvalues.  
By mapping the physical coordinate domain onto a compact interval and factoring out the asymptotic behavior of the wavefunction, the remaining part becomes regular across the interval, allowing the boundary conditions to be imposed as simple vanishing values.  
With a suitable field redefinition and grid discretization, the differential equation is transformed into a matrix equation.  
The quasinormal frequencies are then obtained as the eigenvalues of a generalized matrix eigenvalue problem, which can be efficiently solved using standard numerical algorithms.  
In the present study, we adopt a recent version of the matrix method, which is implemented in hyperboloidal coordinates~\cite{agr-qnm-hyperboloidal-03, agr-qnm-lq-matrix-12} on a Chebyshev grid~\cite{agr-qnm-lq-matrix-11}. The precision and consistency of the numerical results are ensured by repeating the calculations using different grid sizes.

In Fig.~\ref{fig_PT_det_frequency}, we show the effect of the spatial scale of the metric perturbations, in terms of the wave number $k$.
It is observed that the deformation of the QNM spectrum initiates from the high overtones and propagates towards the low-lying modes.
For small wave numbers, the deviations oscillate around their unperturbed values (governed by Eq.~\eqref{qnm_PT}).
As the spatial scale decreases, the amplitude of such deviation increases rapidly.
Specifically, even though the deviations associated with $k=1.0$ indicated by the empty blue squares are clearly identified in the top-left panel, they become almost invisible when presented in the top-right panel, when compared against the deformation triggered by the wave number $k=1.5$, shown in empty orange diamonds.
The latter again becomes insignificant when compared with the deviation represented by empty black triangles for $k=2.0$ shown in the middle-left panel.
In all three panels, the deviations oscillate around the unperturbed QNM spectrum, with higher overtones experiencing more significant deformations.
Moreover, as the wave number increases further, a bifurcation point emerges, and the resulting QNM spectrum divides into two different branches.
The instability continues to propagate towards the low overtones.
It is intriguing to note that such a phenomenon is reminiscent of what was first observed by Skakala and Visser~\cite{agr-qnm-Poschl-Teller-03, agr-qnm-Poschl-Teller-04}, and it was explored further regarding the spectral instability~\cite{agr-qnm-lq-matrix-12}.
However, these existing studies are in a somewhat different and more restrictive context for ``ultraviolet'' metric perturbations featuring discontinuity.

\begin{figure}[ht]
\centerline{
\hspace{0.3cm}
\includegraphics[height=0.35\textwidth]{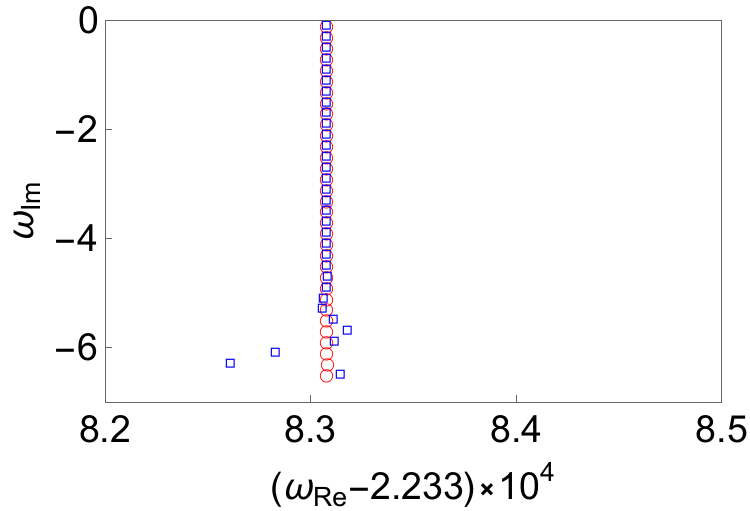}
\hspace{0.2cm}
\includegraphics[height=0.35\textwidth]{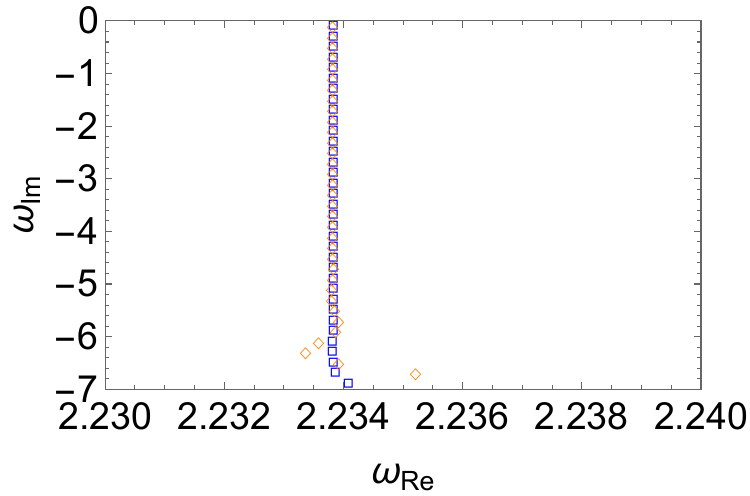}
}
\centerline{
\includegraphics[height=0.35\textwidth]{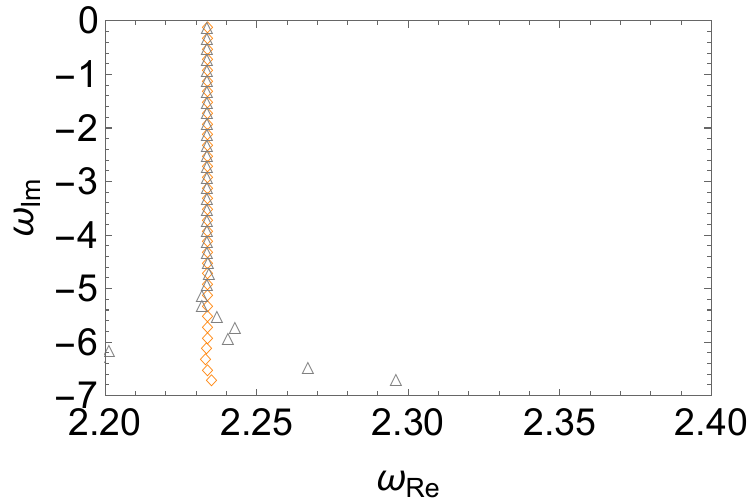}
\includegraphics[height=0.35\textwidth]{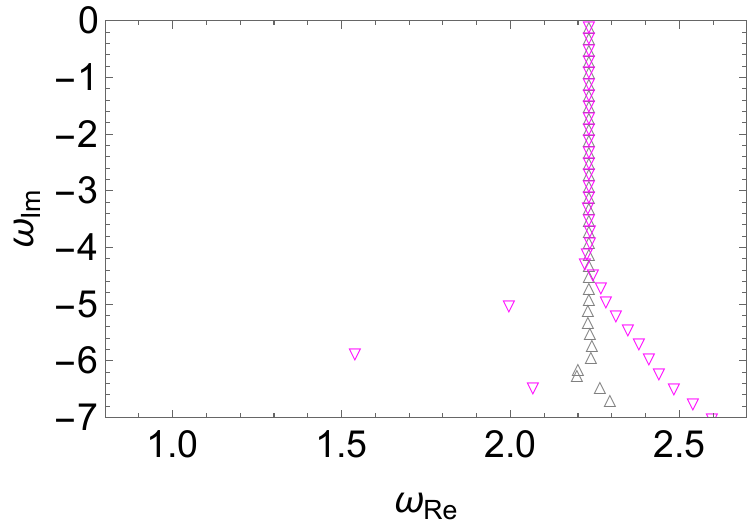}
}
\centerline{
\includegraphics[height=0.35\textwidth]{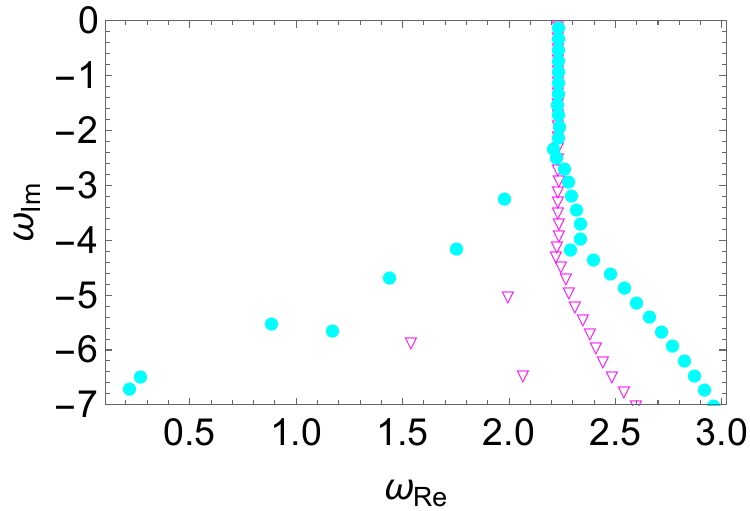}
\includegraphics[height=0.35\textwidth]{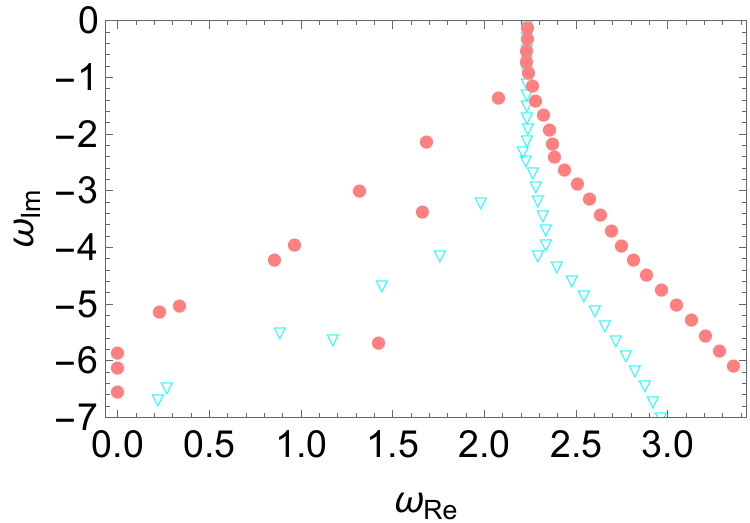}
}
\renewcommand{\figurename}{Fig.}
\vspace{-0.5cm}
\caption{The spectral instability in the perturbed P\"oschl-Teller effective potential Eq.~\eqref{Veff_MPT} triggered by deterministic metric perturbation Eq.~\eqref{V_per_der}, where one assumes $V_0=b=5$ and $k=3.0$.
From left to right and top to bottom, we compare in pairs the resulting QNM spectra for different metric perturbations using the magnitudes $\epsilon=1\times 10^{-8}$ (empty red circles), $1\times 10^{-7}$ (empty blue squares), $1\times 10^{-6}$ (empty orange diamonds), $1\times 10^{-5}$ (empty gray triangles), $1\times 10^{-4}$ (empty magenta flipped triangles), $1\times 10^{-3}$ (filled cyan circles), and $1\times 10^{-2}$ (filled pink squares).
The numerical calculations are carried out using the matrix method, and the results have been verified using different grid sizes.}
\label{fig_PT_det_magnitude}
\end{figure}

In Fig.~\ref{fig_PT_det_magnitude}, we study the effect of the magnitude of the metric perturbations, in terms of the parameter $\epsilon$.
Again, the numerical results obtained by the matrix method have been ascertained using different grid sizes.
The deformation of the QNM spectrum initiates from the high overtones and propagates towards the low-lying modes.
As the magnitude of the perturbation gradually increases and exceeds a particular value (in our specific case, $\epsilon\gtrsim 1\times 10^{-4}$), a bifurcation can be clearly identified.
The bifurcation causes a division of two branches of QNM spectra leaning towards opposite directions.
As the magnitude increases further, the bifurcation propagates towards the low-lying modes, leading to a more significant distortion of the QNM spectrum.
Moreover, as shown in the botton-right panel, the branch which leans towards the imaginary frequency axis is found to turn into purely imaginary modes.
We note that this feature is similar to the imaginary modes observed in~\cite{agr-qnm-lq-matrix-12}.
Different from~\cite{agr-qnm-lq-matrix-12}, where these modes can be confirmed by explicitly evaluating the Wronskian and derived analytically under reasonable approximation, here the results are numerical for the most part, and can only be obtained using an approach based on explicit evaluations of the perturbed effective potential.
In practice, it is known that imaginary modes can be merely a numerical artifact.
Therefore, the claim of their existence (c.f. the special algebraic mode derived by Chandrasekhar~\cite{agr-qnm-18}) must be confirmed with extra caution.
However, based on the strong resemblance between the present results and those derived in~\cite{agr-qnm-lq-matrix-12}, we speculate that the formation of a series of purely imaginary modes in the perturbed P\"oschl-Teller effective potential is a general result.

\begin{figure}[ht]
\centerline{
\hspace{0.3cm}
\includegraphics[height=0.35\textwidth]{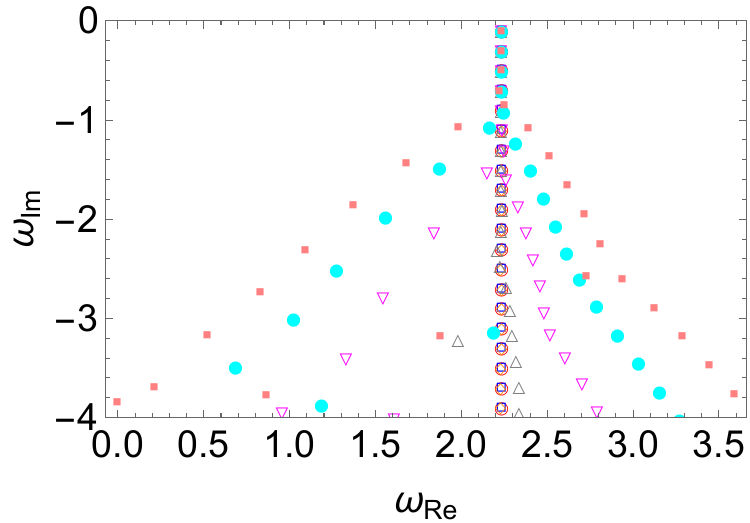}
\hspace{0.2cm}
\includegraphics[height=0.35\textwidth]{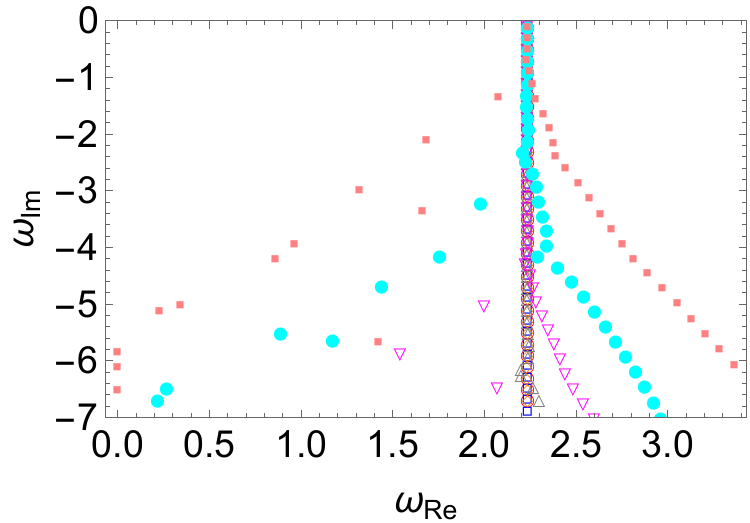}
}
\renewcommand{\figurename}{Fig.}
\vspace{-0.5cm}
\caption{A summary of the deformed QNM spectra shown in Fig.~\ref{fig_PT_det_frequency} (left) and Fig.~\ref{fig_PT_det_magnitude} (right).}
\label{fig_PT_det_sum}
\end{figure}

In Fig.~\ref{fig_PT_det_sum}, we summarize all the spectra presented in Figs.~\ref{fig_PT_det_frequency} and~\ref{fig_PT_det_magnitude}.
From a more general perspective, we see that the spectral instability is more susceptible to more significant, as well as ``ultraviolet'' perturbations, as first pointed out in~\cite{agr-qnm-instability-07}.
A closer analysis indicates that for both large wave numbers and magnitudes, a bifurcation is observed in the spectrum, which propagates from high overtones to the low-lying ones (c.f. Fig.~7 of~\cite{agr-qnm-instability-07}).
For an overtone of a given order, the deformation becomes more significant as the wave number or magnitude of the perturbation increases.
On the other hand, for perturbations of relatively small wave numbers, there is no bifurcation, and the deviation oscillates.
Both features have been observed in the P\"oschl-Teller effective potential~\cite{agr-qnm-lq-matrix-12}, where the metric perturbation was implemented by introducing some discontinuity in terms of a cut or step, and a semi-analytic derivation was feasible.
For the present case, there is no discontinuity, and we speculate that such features are rather general.

\section{Spectral instability triggered by random metric perturbations}\label{sec3}

Now we proceed to analyze random metric perturbations subject to the two following forms
First, we consider the case
\bqn
V_\mathrm{pert} = V_\mathrm{pert}^\mathrm{ran} = x \varepsilon  ,\label{V_per_ran1}
\eqn  
where $\varepsilon$ measures the strength and $x$ is governed by a probability density function $g(r)$, such that $P(r \leq R < r + dr) \approx g(r)\, dr$.
Specifically, we consider the contribution to be a uniform distribution with $-1 \le x \le 1$, giving rise to a uniform white noise satisfying
\bqn
X \sim WN\left(0, \sigma^2=\frac13\right) . \label{GD4UWN}
\eqn
It is noted that the random metric perturbations governed by Eq.~\eqref{GD4UWN} are implemented in a pointwise fashion to ensure the randomness.
Specifically, at individual spatial coordinate points, the magnitude $\epsilon$ will receive a rescaling governed by the variable $X$, whose value is determined entirely independently from those applied to other positions.

\begin{figure}[ht]
\centerline{
\includegraphics[height=0.35\textwidth]{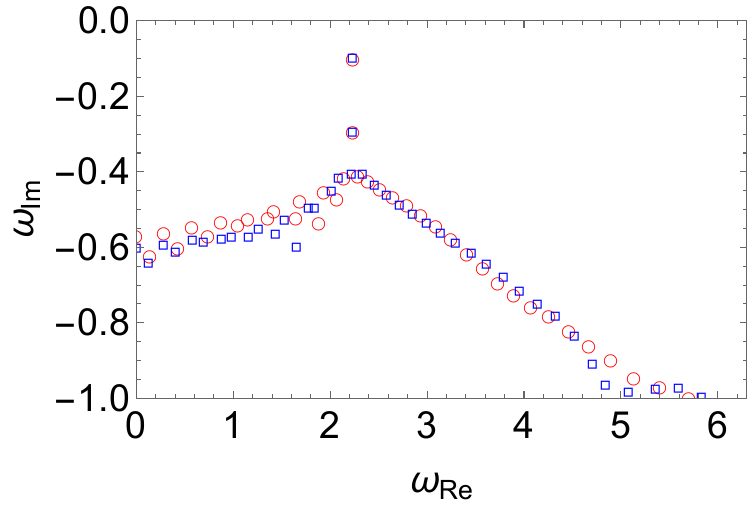}
\includegraphics[height=0.35\textwidth]{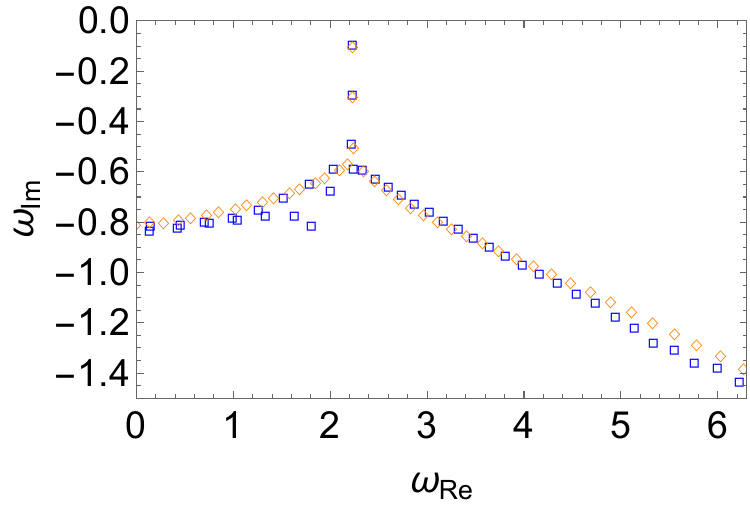}
}
\centerline{
\includegraphics[height=0.35\textwidth]{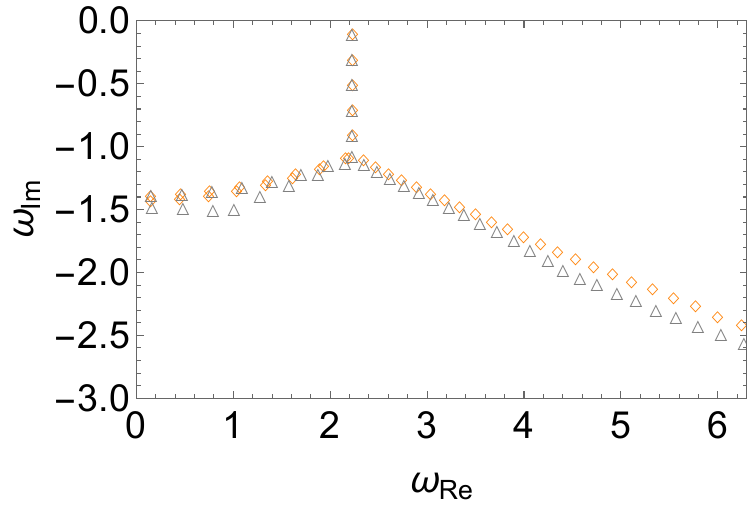}
\includegraphics[height=0.35\textwidth]{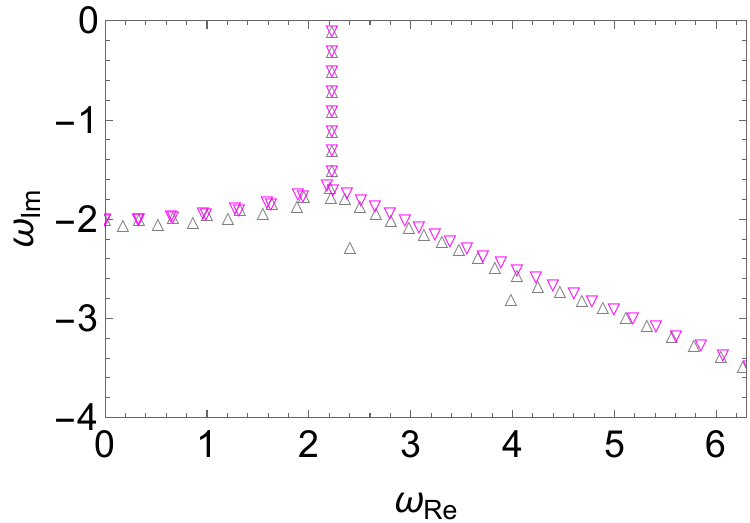}
}
\centerline{
\includegraphics[height=0.35\textwidth]{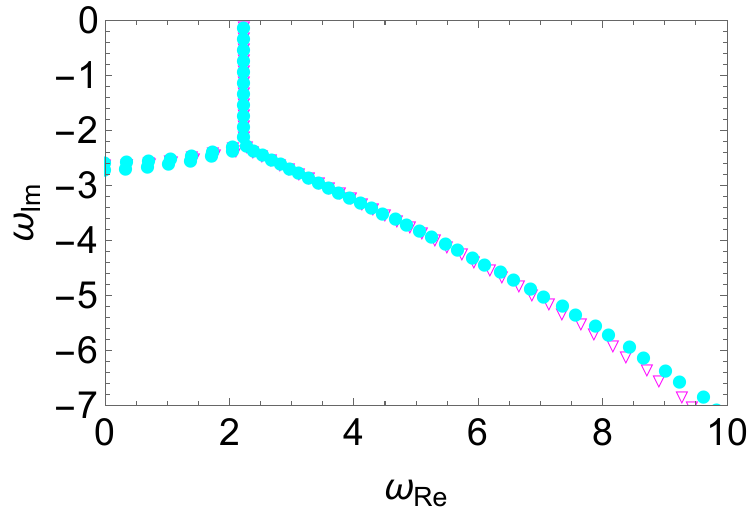}
\includegraphics[height=0.35\textwidth]{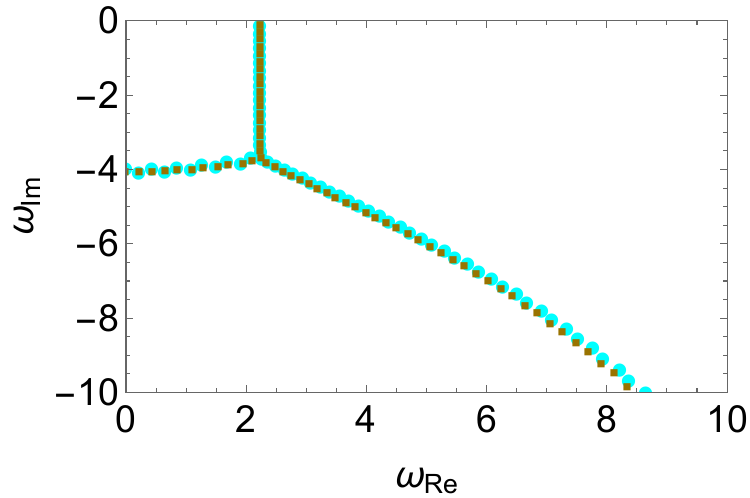}
}
\renewcommand{\figurename}{Fig.}
\vspace{-0.5cm}
\caption{The spectral instability in the perturbed P\"oschl-Teller effective potential Eq.~\eqref{Veff_MPT} triggered by random metric perturbation Eq.~\eqref{V_per_ran1}, where one assumes $V_0=b=5$.
For random metric perturbations of given strength, two sets of results are presented in each panel.
From left to right and top to bottom, we show the resulting QNM spectra for different metric perturbations using the magnitudes $\varepsilon=1\times 10^{-3}$ (empty red circles and empty blue squares), $1\times 10^{-5}$ (empty blue squares and empty orange diamonds), $1\times 10^{-10}$ (empty orange diamonds and empty gray triangles), $1\times 10^{-15}$ (empty gray triangles and empty magneta flipped triangles), $1\times 10^{-20}$ (empty magenta flipped triangles and filled cyan cirlcles), $1\times 10^{-30}$ (filled cyan circles and filled dark-yellow squares).
Due to the nature of random metric perturbations, the perturbed QNM spectra, particularly the high overtones, do not coincide precisely, even though a consistent trend is observed.
The numerical calculations are carried out using the matrix method.}
\label{fig_PT_ran_white}
\end{figure}

The results are shown in Fig.~\ref{fig_PT_ran_white}, where the resulting QNM spectra are evaluated for metric perturbation featuring white noise of different magnitudes.
Owing to the nature of randomness, for the metric perturbation of a given strength, the perturbed QNM spectra, particularly the high overtones, do not coincide.
This is why we present multiple results for given configurations.
Nonetheless, a consistent trend is observed, and in particular, the unperturbed low overtones remain essentially unchanged.
It is somewhat surprising that spectral instability is observed at a minimal level of ``random noise'' in the effective potential.
As shown in the bottom-right panel of Fig.~\ref{fig_PT_ran_white}, even for the magnitude $\varepsilon= 1\times 10^{-30}$, a bifurcation is manifestly shown around $\mathrm{Im}\omega_n\simeq -4$, and a significant deformation in the spectrum is observed for higher overtones.
As the strength of the perturbation increases, the deformation propagates from high overtones toward the low-lying ones.

Secondly, we explore an interplay between random and deterministic perturbations by considering the case
\bqn
V_\mathrm{pert} = V_\mathrm{pert}^\mathrm{det}+V_\mathrm{pert}^\mathrm{ran} ,\label{V_per_ran2}
\eqn
where the random perturbation Eq.~\eqref{V_per_ran1} is added on top of the deterministic metric perturbation Eq.~\eqref{V_per_der}.

\begin{figure}[ht]
\centerline{
\includegraphics[height=0.35\textwidth]{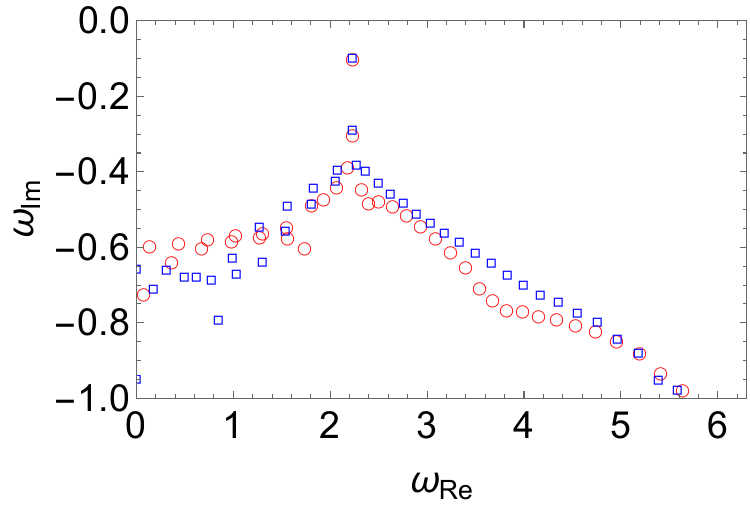}
\includegraphics[height=0.35\textwidth]{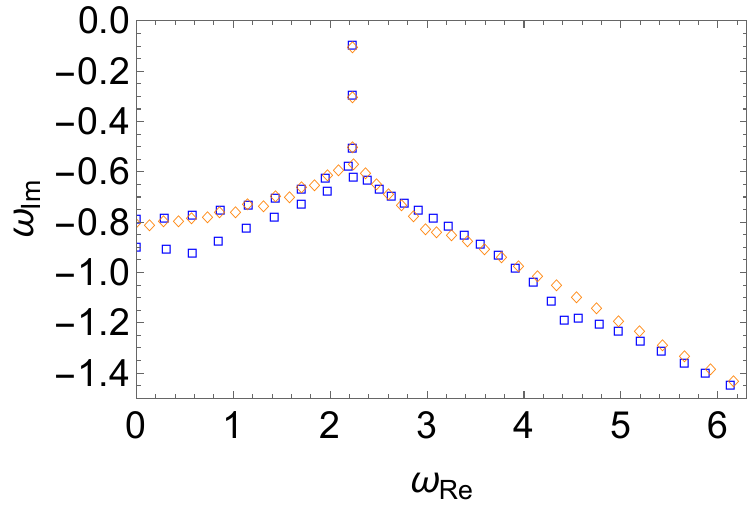}
}
\centerline{
\includegraphics[height=0.35\textwidth]{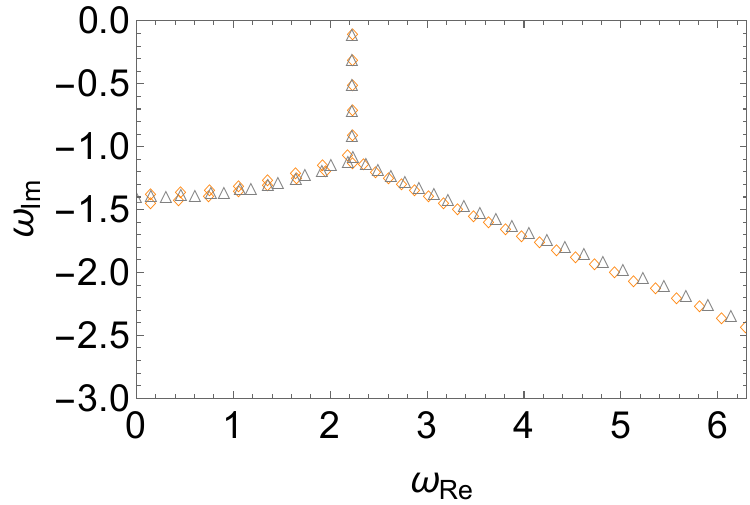}
\includegraphics[height=0.35\textwidth]{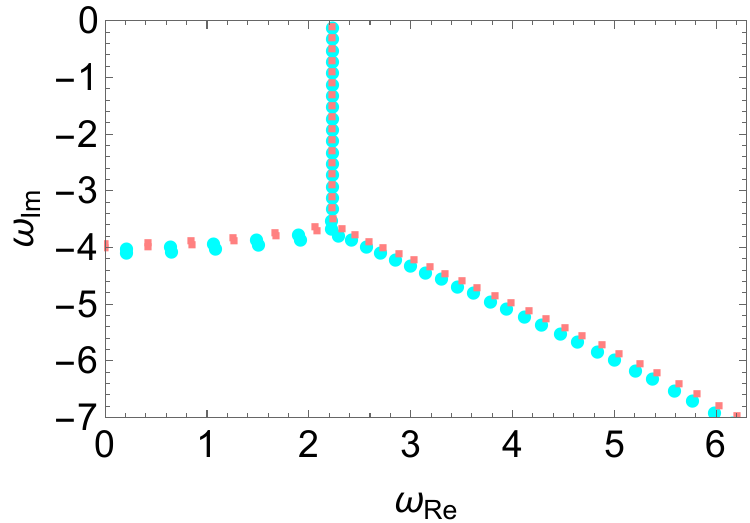}
}
\centerline{
\includegraphics[height=0.35\textwidth]{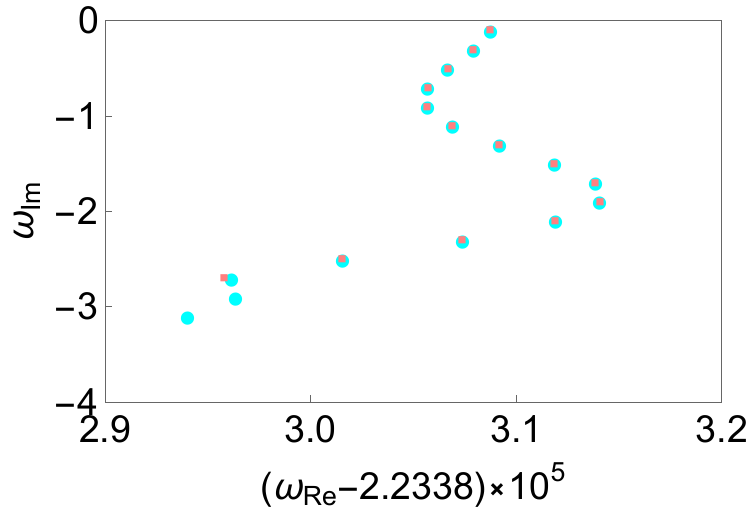}
\includegraphics[height=0.35\textwidth]{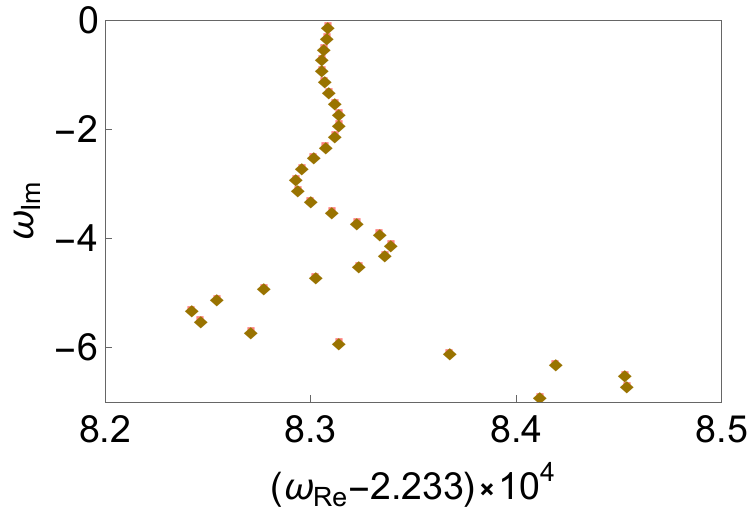}
}
\renewcommand{\figurename}{Fig.}
\vspace{-0.5cm}
\caption{The spectral instability in the perturbed P\"oschl-Teller effective potential Eq.~\eqref{Veff_MPT} triggered by random metric perturbation Eq.~\eqref{V_per_ran2}, where one assumes $V_0=b=5$ and for the deterministic perturbation $k=1$ with strength $\epsilon=1\times 10^{-3}$.
Similar to Fig.~\ref{fig_PT_ran_white}, for random metric perturbations of given strength, two sets of results are presented in each panel.
From left to right and top to bottom, we show the resulting QNM spectra for different metric perturbations using the magnitudes $\varepsilon=1\times 10^{-3}$ (empty red circles and empty blue squares), $1\times 10^{-5}$ (empty blue squares and empty orange diamonds), $1\times 10^{-10}$ (empty orange diamonds and empty gray triangles), $1\times 10^{-30}$ (filled cyan circles and filled pink squares), a magnified view of the slightly perturbed low-lying modes of the last panel with identical $\varepsilon=1\times 10^{-30}$ (filled cyan circles and filled pink squares), $1\times 10^{-50}$ (filled dark-yellow diamonds and filled pink squares).
The numerical calculations are carried out using the matrix method.}
\label{fig_PT_ran_magnitude}
\end{figure}

The numerical results are presented in Fig.~\ref{fig_PT_ran_magnitude}.
As discussed before, for deterministic metric perturbation with a relatively small wave number, one finds that the effect on the low-lying QNMs is perturbative, as the deviation oscillates around those values of the unperturbed metric.
Here, one observes a competition between the two effects.
As shown in the upper and middle rows, as the magnitude of random metric perturbation increases, a bifurcation point appears that subsequently propagates toward the low-lying modes.
The uncertainty of the perturbed QNMs is attributed to the randomness, which becomes more pronounced as the magnitude $\varepsilon$ increases.
On the other hand, as the randomness becomes sufficiently small, around $\varepsilon\simeq 1\times 10^{-30}$, the oscillating deviation in the low-lying modes becomes apparent, as demonstrated in the bottom row of Fig.~\ref{fig_PT_ran_magnitude}.
In particular, as one compares the two panels in the bottom row, the oscillation becomes more significant as the random metric perturbation decreases.

\begin{figure}[ht]
\centerline{
\hspace{0.3cm}
\includegraphics[height=0.35\textwidth]{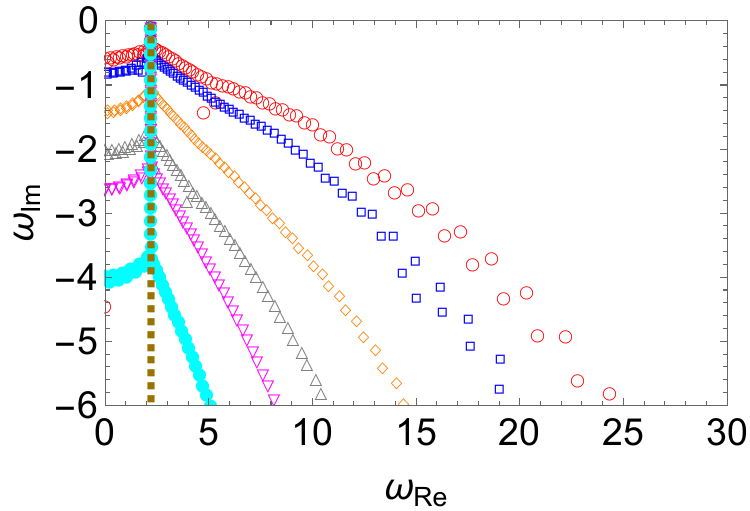}
\hspace{0.2cm}
\includegraphics[height=0.35\textwidth]{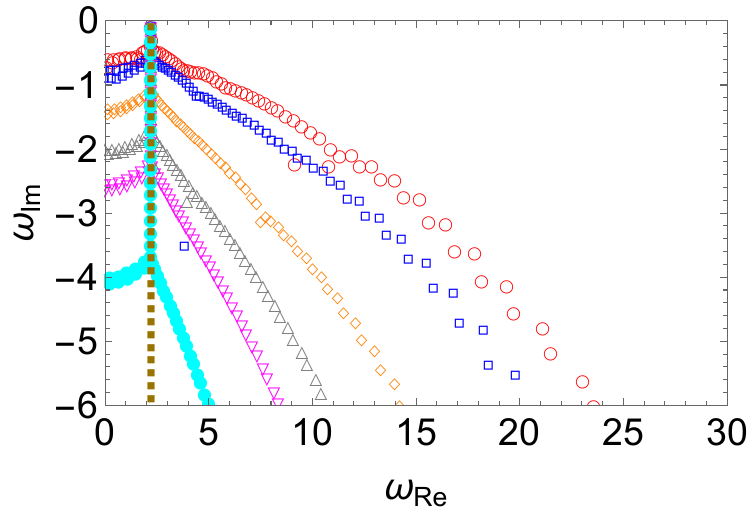}
}
\renewcommand{\figurename}{Fig.}
\vspace{-0.5cm}
\caption{A summary of the deformed QNM spectra shown in Fig.~\ref{fig_PT_ran_white} (left) and Fig.~\ref{fig_PT_ran_magnitude} (right), where one enlists the following strengths of the random perturbations $\epsilon=1\times 10^{-3}$, $1\times 10^{-5}$, $1\times 10^{-10}$, $1\times 10^{-15}$, $1\times 10^{-20}$, $1\times 10^{-30}$, $1\times 10^{-50}$.}
\label{fig_PT_ran_sum}
\end{figure}

\begin{figure}[ht]
\centerline{
\includegraphics[height=0.35\textwidth]{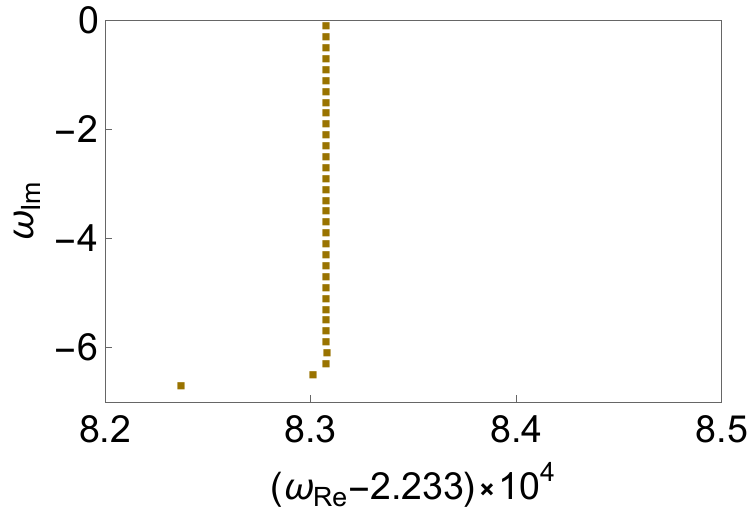}
}
\renewcommand{\figurename}{Fig.}
\vspace{-0.5cm}
\caption{A magnified view of the low-lying modes of the random metric perturbation Eq.~\eqref{V_per_ran1} with $\varepsilon=1\times 10^{-50}$ (similar to the panels shown in Fig.~\ref{fig_PT_ran_white}), as a direct comparison with the bottom-right panel of Fig.~\ref{fig_PT_ran_magnitude}.}
\label{fig_PT_ran_white2}
\end{figure}

In Fig.~\ref{fig_PT_ran_sum}, we perform a comparison between the deformed QNM spectra presented in Figs.~\ref{fig_PT_ran_white} and~\ref{fig_PT_ran_magnitude}.
Overall, one immediately identifies a strong resemblance between the two panels.
By comparing the deviation of a given overtone of the spectrum triggered by random perturbations of different magnitudes, the spectral instability is more pronounced as the magnitude of the perturbation increases.
However, while observing more closely, one might wonder whether the oscillating deviation observed in the bottom-right panel of Fig.~\ref{fig_PT_ran_magnitude} might also appear in the case of random metric perturbations consisting of pure white noise.
This is not the case, as manifestly demonstrated in Fig.~\ref{fig_PT_ran_white2}.

\section{Sensitivity of spectral instability on the localization of the perturbation}\label{sec4}

Last but not least, we investigate the effect of the location of the metric perturbation.
As shown in Refs.~\cite{agr-qnm-instability-15, agr-qnm-instability-58, agr-qnm-instability-56, agr-qnm-lq-matrix-12}, the spectral instability becomes more severe as the metric perturbation moves toward spatial infinity.
However, as pointed out in~\cite{agr-qnm-instability-16}, to address this point more clearly, one must carefully consider metric perturbations that are physically relevant.

For this purpose, we consider the following localized perturbation of Gaussian shape in tortoise coordinates
\bqn
V_\mathrm{pert}^\mathrm{loc} =  \frac{\epsilon}{\sigma\sqrt{2\pi}} e^{-\frac12\left(\frac{(r_*-r_c)}{\sigma}\right)^2} ,\label{VperLoc}
\eqn
with strength $\epsilon$, located at $r_*=r_c$ with standard deviation $\sigma$.
Specifically, the magnitude $\epsilon$ is scaled so that $\epsilon(r_c=r_0)=\epsilon_0$, where $r_0$ is the initial location of the perturbation.
Here, we choose a particular scheme where $\epsilon$ is inversely proportional to the squared distance
\begin{equation}
\epsilon = \epsilon_0 \left(\frac{r_0}{r_c}\right)^2 .\label{epsilonScale}
\end{equation}
This is motivated to simulate a perturbation whose total energy roughly remains constant as its location moves away from the compact object.

\begin{figure}[ht]
\centerline{
\hspace{0.3cm}
\includegraphics[height=0.35\textwidth]{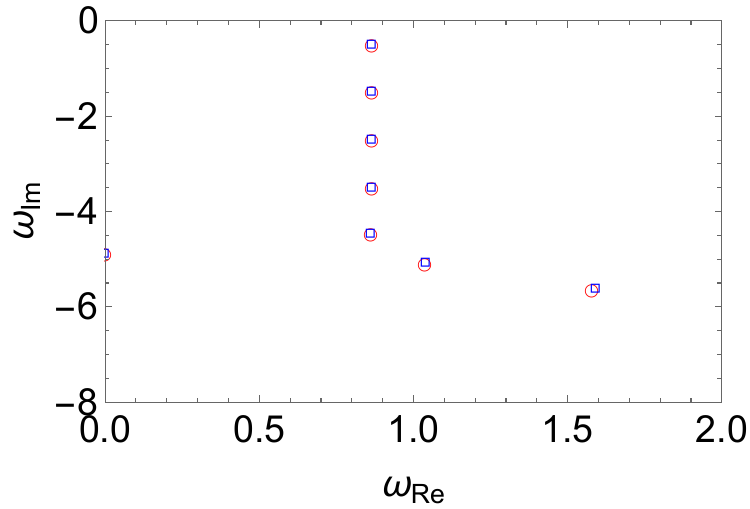}
\hspace{0.2cm}
\includegraphics[height=0.35\textwidth]{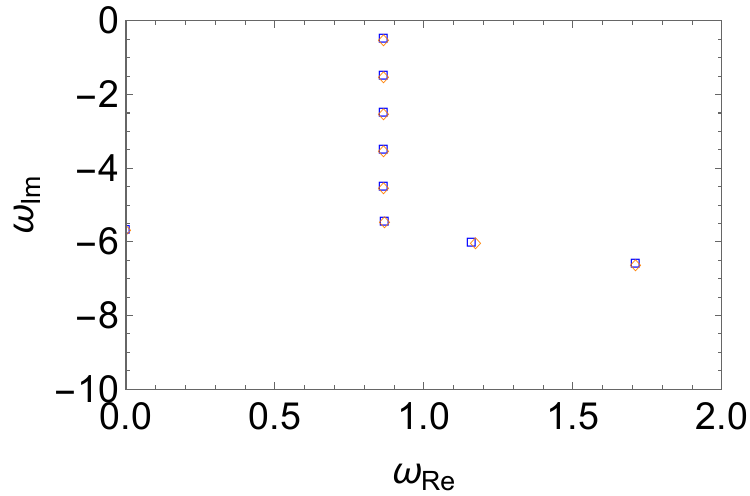}
}
\centerline{
\includegraphics[height=0.35\textwidth]{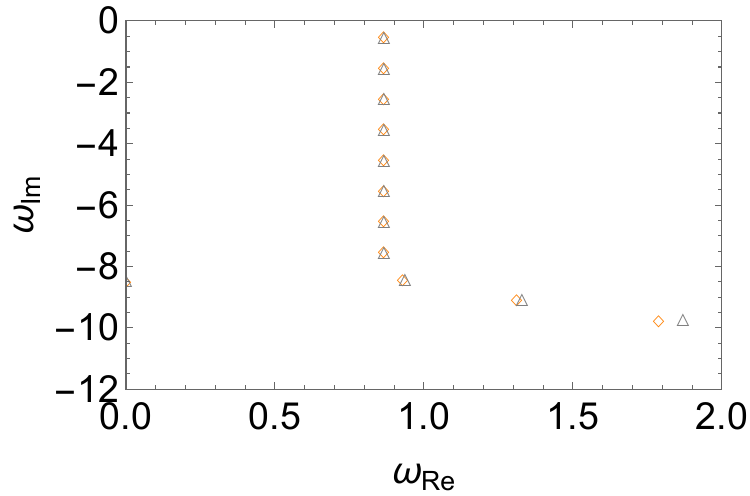}
\includegraphics[height=0.35\textwidth]{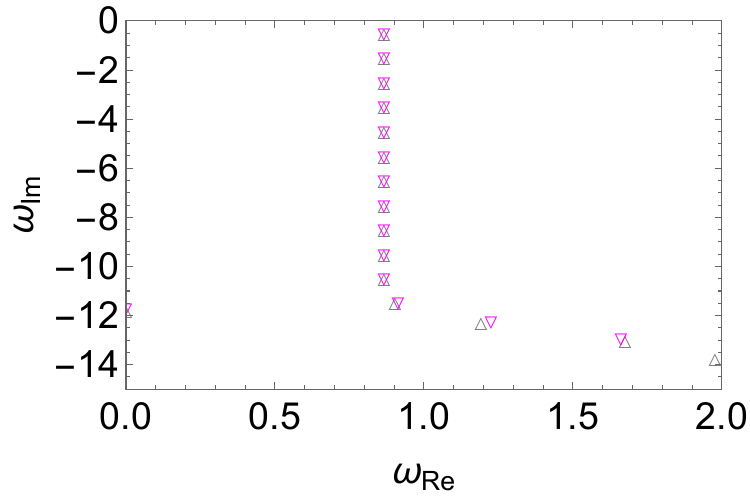}
}
\centerline{
\includegraphics[height=0.35\textwidth]{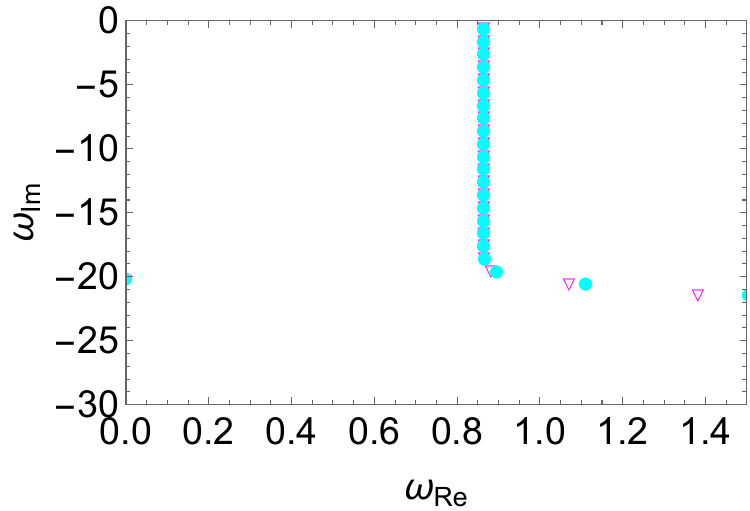}
\includegraphics[height=0.35\textwidth]{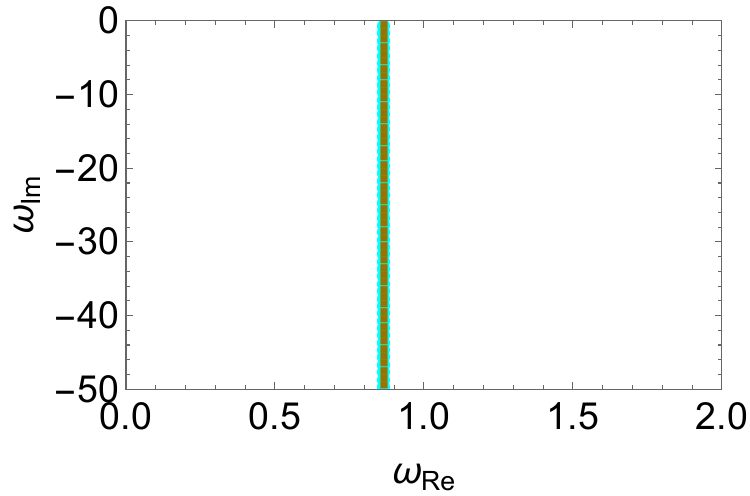}
}
\renewcommand{\figurename}{Fig.}
\vspace{-0.5cm}
\caption{The spectral instability in the perturbed P\"oschl-Teller effective potential Eq.~\eqref{Veff_MPT} triggered by deterministic metric perturbation Eq.~\eqref{VperLoc} placed at different radial coordinates, where one assumes $V_0=b=1$, $\epsilon_0=1\times 10^{-10}$, $\sigma=1$, and $r_0=2$.
For a given metric perturbation, two sets of results are presented in each panel.
From left to right and top to bottom, we compare in pairs the resulting QNM spectra for different metric perturbations using different locations $r_c=5$ (empty red circles and empty blue squares), $7$ (empty blue squares and empty orange diamonds), $10$ (empty orange diamonds and empty gray triangles), $12$ (empty gray triangles and empty magneta flipped triangles), $15$ (empty magenta flipped triangles and filled cyan cirlcles), $20$ (filled cyan circles and filled dark-yellow squares).
The numerical calculations are carried out using the matrix method.}
\label{fig_PT_det_location}
\end{figure}

The numerical results are presented in Fig.~\ref{fig_PT_det_location}.
Owing to the scaling Eq.~\eqref{epsilonScale}, one observes that the spectral instability becomes less pronounced as the location of the perturbation moves away.
Apparently, such a result is in agreement with physical intuition that a given metric perturbation should introduce less of an impact to the gravitational system when it is planted further away from it.
Meanwhile, it is in contrast to most existing studies, where an opposite effect has been observed in not only the high overtones~\cite{agr-qnm-lq-matrix-12} but also the fundamental mode~\cite{agr-qnm-instability-15, agr-qnm-instability-58, agr-qnm-instability-56}.
The primary reason for this difference is that those studies considered a more straightforward setup in which the magnitude of the perturbation remains unchanged.
Therefore, the present result indicates that the physical relevance of the perturbation plays a crucial role in such analyses.
We note that in Fig.~\ref{fig_PT_det_location}, we take $V_0=1$ and $b=1$ to make the perturbation more pronounced.
The numerical results indicate that the spectral instability becomes less evident as the perturbation moves away, even for such a choice.
Nonetheless, owing to the direct impact of the fundamental mode on the time-domain waveforms, it is arguable that the observational implications are substantial.

\section{Concluding remarks}\label{sec5}

To summarize, we analyzed the spectral instability triggered by random and deterministic metric perturbations.
We investigate the dependence on the magnitude, spatial scale, and localization of the metric perturbations, and discuss the underlying physical interpretations.
Small metric perturbations were observed to initially have a limited impact on the less damped black hole quasinormal modes, and deviations typically oscillate around their unperturbed values, in agreement with a phenomenon first derived by Skakala and Visser in a more restrictive context.
As either the frequency or the magnitude of the perturbations increases, in the higher overtone region, the deformation propagates, amplifies, and eventually gives rise to spectral instability and, inclusively, bifurcation in the quasinormal mode spectrum. 
On the one hand, deterministic metric perturbations yield a deformed but well-defined quasinormal spectrum.
On the other hand, random perturbations lead to uncertainties in the resulting spectrum, while the primary trend of the spectral instability shows consistency.
Additionally, the location of the perturbation was found to play a significant role in the resulting quasinormal mode spectrum.
In contrast to most results in the literature, we show that the observed spectral instability might become suppressed as the metric perturbation moves away from the compact object, once the underlying perturbations are devised to be physically appropriate.  
We conclude by pointing out substantial observational implications of the spectral instability, and plan to continue to work on this pertinent topic.

\section*{Acknowledgements}

We gratefully acknowledge the financial support from Brazilian agencies 
Funda\c{c}\~ao de Amparo \`a Pesquisa do Estado de S\~ao Paulo (FAPESP), 
Funda\c{c}\~ao de Amparo \`a Pesquisa do Estado do Rio de Janeiro (FAPERJ), 
Conselho Nacional de Desenvolvimento Cient\'{\i}fico e Tecnol\'ogico (CNPq), 
and Coordena\c{c}\~ao de Aperfei\c{c}oamento de Pessoal de N\'ivel Superior (CAPES).
This work is supported by the National Natural Science Foundation of China (NSFC).
A part of this work was developed under the project Institutos Nacionais de Ci\^{e}ncias e Tecnologia - F\'isica Nuclear e Aplica\c{c}\~{o}es (INCT/FNA) Proc. No. 464898/2014-5.
This research is also supported by the Center for Scientific Computing (NCC/GridUNESP) of São Paulo State University (UNESP).

\bibliographystyle{h-physrev}
\bibliography{references_qian}

\end{document}